\documentclass[runningheads,11points]{llncs}

\usepackage{amssymb,amstext}
\usepackage{cite,scrtime}
\usepackage[bookmarks=false]{hyperref}
\usepackage{graphics}
\usepackage{amsmath,dsfont}
\usepackage{mathrsfs}
\usepackage{fancyhdr}
\usepackage{verbatim}
\usepackage{boxedminipage}
\usepackage{colortbl}
\usepackage{algorithm}
\usepackage[noend]{algorithmic}
\usepackage{tikz}

\usepackage[T1]{fontenc}
\usepackage{tikz}
\usepackage{graphicx}

\newcommand{\Rcal}{\mathcal{R}}

\newcommand{\Ocal}{\mathcal{O}}

\newcommand{\TT}{\mathfrak{T}}

\newtheorem{invariant}{Invariant}

\newcommand{\sm}{\setminus}

\newenvironment{tight_enumerate}{
\begin{enumerate}
  \setlength{\itemsep}{2pt}
  \setlength{\parskip}{0pt}
}{\end{enumerate}}

\newcommand{\tw}{{\mathbf{tw}}}

\newcommand{\dotsss}[1]{[#1]}

\newcommand{\es}{\emptyset}

\newcommand{\ps}{{\sf ps}}

\newcommand{\outb}[1]{\draw[dotted] (#1) -- +(0.5, 0.5);\draw[dotted] (#1) -- +(-0.5, 0.5);}
\newcommand{\capt}[2]{\draw[white] (#1) -- node[above,sloped,black] {#2} +(2,0)}

\tikzstyle{vertex}=[
  circle,
  draw,
  thick,
  inner sep=0.1cm,
  minimum size=2mm
]

\title{Efficient FPT algorithms for (strict) compatibility of unrooted phylogenetic trees\thanks{An extended abstract of this work will appear in the \emph{Proceedings of the 11th  International Conference on Algorithmic Aspects of Information and Management (\textbf{AAIM}), Bergamo, Italy, July \textbf{2016}}.}
}

\author{Julien Baste~\inst{1}  \and Christophe Paul~\inst{1} \and Ignasi Sau~\inst{1} \and Celine Scornavacca~\inst{2}}

\authorrunning{Julien Baste, Christophe Paul, Ignasi Sau, and Celine Scornavacca}
\titlerunning{ FPT algorithms for compatibility of unrooted phylogenetic trees}

\institute{CNRS, LIRMM, Universit\'e de Montpellier, Montpellier, France.\\
\email{\{baste, paul, sau\}@lirmm.fr} \and
Institut des Sciences de l'Evolution (Université de Montpellier, CNRS, IRD, EPHE), Montpellier, France.\\
\email{celine.scornavacca@umontpellier.fr}}

\begin{document}
\maketitle

\begin{abstract}

In phylogenetics, a central problem is to infer the evolutionary relationships between a set of species $X$; these relationships are often depicted via a phylogenetic tree -- a tree having its leaves univocally labeled by elements of $X$ and without degree-2 nodes --  called the ``species tree''.
One common approach for reconstructing a species tree consists in first constructing several phylogenetic trees from primary data (e.g. DNA sequences originating from some species in $X$), and then constructing a single phylogenetic tree maximizing the ``concordance'' with the input trees. The so-obtained tree is our estimation of the species tree and, when the input trees are defined  on overlapping -- but not identical -- sets of labels,  is called ``supertree''. In this paper, we focus on two problems that are central when combining  phylogenetic trees into a supertree: the compatibility and the strict compatibility problems for unrooted phylogenetic trees.  These problems are strongly related, respectively, to the notions of ``containing as a minor'' and ``containing as a topological minor'' in the graph community. Both problems are known to be  fixed-parameter tractable in the number of input trees $k$, by using their expressibility in Monadic Second Order Logic and a reduction to graphs of bounded treewidth. Motivated by the fact that the dependency on $k$ of these algorithms is prohibitively large, we give the first explicit dynamic programming algorithms for solving these problems, both running  in time $2^{O(k^2)} \cdot n$, where $n$ is the total size of the input.


\vspace{.2cm}

\textbf{Keywords}: Phylogenetics; compatibility; unrooted phylogenetic trees; parameterized complexity; FPT algorithm; dynamic programming.  
\end{abstract}

\section{Introduction}
\label{sec:intro}


A central goal in \emph{phylogenetics} is to
clarify the relationships of extant species in an evolutionary context. 
Evolutionary relationships are commonly represented via  \emph{phylogenetic trees}, that is,  acyclic connected graphs where leaves are univocally labeled by a label set $X$, and without degree-2 nodes. 
When a phylogenetic tree is defined on a label set $X$ designating a set of genes issued from a gene family, we refer to it as a \emph{gene tree}, while, when $X$ corresponds to a set of  extant species, we refer to it as a \emph{species tree}.
A gene tree can differ from the species tree depicting the evolution of the species containing the gene for a number of reasons  \cite{maddison1989reconstructing}. Thus, a common way to estimate a species tree for a set of species $X$ is to choose \emph{several} gene families  that  appear in the genome of the  species in $X$, reconstruct a gene tree per each gene family (see \cite{felsenstein2004inferring} for a detailed review of how to infer phylogenetic trees), and finally combine the trees in a unique tree that maximizes the ``concordance'' with the given gene trees.
The rationale  underlying this approach is the confidence that, using several genes,
the species signal will prevail and emerge from the conflicting gene trees.
If the gene trees are all defined on the same label set, we are in the \emph{consensus} setting; otherwise the  trees are defined on overlapping -- but not identical -- sets of labels, and we are in the \emph{supertree} setting. Several consensus and supertree methods exist in the literature (see \cite{scornavacca2009supertree,bininda2004phylogenetic,bininda2002super} for a review), and they differ in the way the concordance is defined.

In this paper, we focus on a problem that arises in the supertree setting: given a set of gene trees $\mathcal{T}=\{T_1, \ldots, T_k\}$ on label sets $\{X_1, \ldots, X_k\}$, respectively, does there exist a species tree on $X:=\cup_{i=1}^k X_i$ that \emph{displays} all the trees in  $\mathcal{T}$? This is the so-called {\sc Compatibility of Unrooted Phylogenetic Trees} problem.   The notion of ``displaying'' used by the phylogenetic community, which will be formally defined in Section~\ref{sec:defs},  coincides with that of ``containing as a minor'' in the graph community. Another related problem is the  {\sc Strict Compatibility} (or {\sc Agreement}) {\sc of Unrooted Phylogenetic Trees} problem, where the notion of ``displaying'' is replaced by that of ``strictly displaying''.  This notion, again defined formally in Section~\ref{sec:defs}, coincides with that of ``containing as a topological minor'' in the graph community.

Both problems are polynomial-time solvable when the given gene trees are out-branching (or {\emph{rooted} in the phylogenetic literature), or all contain some common label  \cite{Aho81,oneTreeStrict1996}. In the general case, both problems are NP-complete \cite{Steel92} and fixed-parameter tractable in the number of trees $k$ \cite{compatibility2006,scornavacca2014agreement}.
The fixed-parameter tractability of these problems has been established via Monadic Second Order Logic (MSOL)  together with a reduction to graphs of bounded treewidth. For both problems, it can be checked that the corresponding MSOL formulas~\cite{compatibility2006,scornavacca2014agreement} contain 4 alternate quantifiers, implying by~\cite{FrickG04} that the dependency on $k$ in the derived algorithms is given by a  tower
of exponentials of  height 4; clearly, this is prohibitively large for practical applications. Therefore, even if the notion of compatibility has been defined quite some time ago~\cite{gordon1986consensus}, at the moment no ``reasonable'' FPT algorithms exist for these problems, that is, algorithms with running time $f(k) \cdot p(|X|)$, with $f$ a moderately growing function  and $p$ a low-degree polynomial. 
In this paper we fill this lack and we prove the following two theorems.

\begin{theorem}\label{thm:minor}
The {\sc Compatibility of Unrooted Phylogenetic Trees} problem can be solved in time $2^{O(k^2)} \cdot n$, where $k$ is the number of trees and $n$ is the total size of the input.
\end{theorem}

\begin{theorem}\label{thm:topo}
The {\sc Agreement of Unrooted Phylogenetic Trees} problem can be solved in time $2^{O(k^2)} \cdot n$,  where $k$ is the number of trees and $n$ is the total size of the input.
\end{theorem}

Our approach for proving the two above theorems is to present explicit dynamic programming algorithms on graphs of bounded treewidth. As one could suspect from the fact that the corresponding MSOL formulas are quite involved~\cite{compatibility2006,scornavacca2014agreement}, it turns out that our dynamic programming algorithms are quite involved as well, implying that we are required to use a technical data structure.

This paper is organized as follows.  In Section~\ref{sec:defs} we provide some preliminaries and we define the problems under study. In Section~\ref{sec:minor}  we present our algorithm for the {\sc Compatibility of Unrooted Phylogenetic Trees} problem, and the algorithm for the {\sc Agreement of Unrooted Phylogenetic Trees} problem is given in Section~\ref{sec:topological}. Finally, we provide some directions for further research in Section~\ref{sec:conclusions}.



\section{Preliminaries}
\label{sec:defs}








\paragraph{\textbf{\emph{Basic definitions}}.} Given a positive integer $k$, we denote by $[k]$ the set of all integers between $1$ and $k$.
{If $S$ is a set, we denote by $2^S$ the set of all subsets of $S$.} A \emph{tree} $T$ is an acyclic connected graph. We denote by $V(T)$ its vertex set, by $E(T)$ its edge set, and by  $L(T)$ its set of vertices of degree one, called \emph{leaves}. Two trees $T$ and $T'$ are \emph{isomorphic} if there is a bijective function $\alpha:V(T)\cup E(T)\to V(T')\cup E(T')$ such that for every edge $e=\{u,v\}\in E(T)$, $\alpha(e)=\{\alpha(u),\alpha(v)\}$.
{If $T$ is a tree and $S$ is a subset of $V(T)$, we denote by $T[S]$ the subgraph of $T$ induced by $S$.}
\emph{Suppressing} a degree-2 vertex $v$ in a graph $G$ consists in deleting $v$ and adding an edge between the former neighbors  of $v$, if they are not already adjacent.
\emph{Identifying} two vertices $v$ and $v'$ of a graph $G$  consists in creating a graph
$H$ by
{removing $v$ and $v'$ and}
adding a new vertex
{$w$ such that, for each $u \in V(G) \sm \{v,v'\}$, there is an edge $\{u,w\}$ in $E(H)$ if and only if $\{u,v\} \in E(G)$ or $\{u,v'\} \in E(G)$}.
\emph{Contracting} an edge $e=\{u,v\}$ in $G$ consists in
identifying $u$ and $v$. A graph $H$ is a \emph{minor} (resp. \emph{topological minor}) of a graph $G$ if $H$ can be obtained from a subgraph of $G$ by contracting edges (resp. contracting edges with at least one vertex of degree $2$). See~\cite{Die10} for more details about the notions of minor and topological minor.
If $Y$ is a subset of vertices of a tree $T$, then $T|_Y$ is the tree obtained from the minimal subtree of $T$ containing $Y$ by suppressing degree-$2$ vertices. For simplicity, we may sometimes consider the vertices of $T|_Y$ also as vertices of $T$.

As already mentioned in the introduction,  an \emph{unrooted phylogenetic tree}  on a label set $X$ is defined as a pair $(T, \phi)$ with $T$ a
tree with no degree-$2$ vertex along with  a  bijective function $\phi: L(T) \rightarrow X$.
{We say that a vertex $v \in L(T)$ is \emph{labeled} with label $\phi(v)$.}
Two unrooted phylogenetic trees $(T, \phi)$ and $(T', \phi')$ are \emph{isomorphic} if there exists an isomorphism $\alpha$ from $T$ to $T'$ satisfying that if $v \in L(T)$ then $\phi'(\alpha(v)) = \phi(v)$.

The three graph operations defined above, namely suppressing a vertex, identifying two vertices, and contracting an edge,  can be naturally generalized to unrooted phylogenetic trees. In this context, two vertices to be identified are either both unlabeled or both with the same label. In the latter case, the newly created vertex inherits the label of the identified vertices. Finally, contractions in unrooted phylogenetic trees are restricted to edges incident to two unlabeled vertices.
{In this case, we speak about \emph{upt-contraction}.}
If $(T,\phi)$ is an unrooted phylogenetic tree and $Y$ is subset of leaves of $L(T)$, then $(T ,\phi)|_Y$  is the unrooted phylogenetic tree $(T|_Y,\phi|_Y)$ where $\phi|_Y$ is the restriction of $\phi$ to the label set $Y$.

\paragraph{\textbf{\emph{(Strictly) Compatible supertree}}.}
Let $\mathcal{T}=\{(T_1, \phi_1), (T_2, \phi_2), \ldots, (T_k, \phi_k)\}$ be a collection of unrooted phylogenetic trees, not necessarily on the same label set. We say that an unrooted phylogenetic tree $(T,\phi)$ is a  \emph{compatible supertree} of $\mathcal{T}$
 {if for every $i \in [k]$,} $(T_i, \phi_i) \in \mathcal{T}$ can be obtained from $(T,\phi)|_{L(T_i)}$ by
 {performing upt-contractions}.
The phylogenetic tree $(T,\phi)$ is a \emph{strictly compatible supertree} of $\mathcal{T}$ {if for every $i \in [k]$,} $(T_i, \phi_i) \in \mathcal{T}$ is isomorphic to $(T, \phi) |_{L(T_i)}$. If a collection $\mathcal{T}$ of unrooted phylogenetic trees admits a (strictly) compatible supertree, then we say that $\mathcal{T}$ is \emph{(strictly) compatible}. The two definitions are equivalent when $\mathcal{T}$ contains only binary phylogenetic trees, that is, unrooted trees in which every vertex that is not a leaf has degree 3. Note that, as mentioned in the introduction, the notions of ``being a compatible supertree'' and ``being a strictly compatible supertree'' correspond, modulo the conditions on the labels, to the notions of ``containing as a minor'' and ``containing as a topological minor'', respectively.

%
In this paper we consider  the following problem:

\medskip
\noindent
{\sc Compatibility of Unrooted Phylogenetic Trees}\\
Instance: A set $\mathcal{T}$ of $k$ unrooted phylogenetic trees.\\
Parameter: $k$.\\
Question: Does there exist an unrooted phylogenetic tree $(T,\phi)$ {that is a  compatible supertree of $\mathcal{T}$?}



\vspace{.4cm}

The {\sc Agreement (or Strict Compatibility) of Unrooted Phylogenetic Trees} problem is defined analogously, just by replacing ``compatible supertree'' with ``strictly compatible supertree''.
%
For notational simplicity,  we may henceforth drop the function $\phi$ from an unrooted phylogenetic tree $(T,\phi)$, and just assume that each leaf of $T$ comes equipped with a label.


Assume that $\widehat{T}$ is a compatible supertree of $\mathcal{T}$. Then, according to the definition of minor, for every $i\in [k]$, every vertex $v\in V(T_i)$ can be mapped to a subtree of $\widehat{T}$, in such a way that the subtrees corresponding to the vertices of the same tree are pairwise disjoint. We call the set of vertices of that subtree the \emph{vertex-model} of $v$. Observe that by the definition of the {upt-contraction} operation, the vertex-model of a leaf is a singleton. Hereafter, we denote by $\widehat{\varphi}(v)$ the subset of vertices belonging to the vertex-model of $v$. Moreover, if $u,v \in V(T_i)$ are two adjacent vertices in $T_i$, then there is {\sl exactly one} edge in $\widehat{T}$ that connects the vertex-model of $u$ to the vertex-model of $v$. We call such an edge of $\widehat{T}$ the \emph{edge-model} of  $\{u,v\} \in E(T_i)$. Observe that a vertex of $\widehat{T}$ may belong to several vertex-models, but then these vertex-models correspond to vertices from different trees of $\mathcal{T}$. Also, an edge of $\widehat{T}$ may be the edge-model of edges of different trees of $\mathcal{T}$.

Similarly, if $\widehat{T}$  is a strictly compatible supertree of $\mathcal{T}$, then according to the definition of topological minor, for every $i\in [k]$, every vertex $v\in V(T_i)$ can be mapped to a vertex of $\widehat{T}$, called the \emph{vertex-model} of $v$,  in such a way that this mapping is injective when restricted to every $i\in [k]$. In this case,  if $u,v \in V(T_i)$ are two adjacent vertices in $T_i$, then there is exactly one {\sl path} in $\widehat{T}$ that connects the vertex-model of $u$ to the vertex-model of $v$ called the \emph{edge-model} of  $\{u,v\} \in E(T_i)$. Similarly to the vertex-models, the edge-models of the same tree need to be pairwise disjoint, except possibly for their endvertices.

\paragraph{\textbf{\emph{Treewidth}}.}
A \emph{tree-decomposition} of width $w$ of a graph $G=(V,E)$ is a pair $({\sf T}, \mathcal{B})$, where ${\sf T}$ is a tree and $\mathcal{B} = \{ B_t \mid B_t \subseteq V, t \in V({\sf T}) \}$ such that
\begin{itemize}
\item[$\bullet$] $\bigcup_{t \in V({\sf T})} B_t = V$,
\item[$\bullet$] for every edge $\{u,v\} \in E$ there is a $t \in V({\sf T})$ such that $\{u, v\} \subseteq B_t$,
\item[$\bullet$] $B_i \cap B_k \subseteq B_j$ for all $\{i,j,k\} \subseteq V({\sf T})$ such that $j$ lies on the unique path from $i$ to $k$ in ${\sf T}$, and
\item[$\bullet$] $\max_{t \in V({\sf T})} |B_t| = w +1$.
\end{itemize}

To avoid confusion, we speak about the \emph{nodes} of a tree-decomposition and the \emph{vertices} of a graph. The sets of $\mathcal{B}$ are called \emph{bags}. The \emph{treewidth} of $G$, denoted by $\tw(G)$, is the smallest integer $w$ such that there is a tree-decomposition of $G$ of width $w$.


\begin{theorem}[Bodlander \emph{et al.} \cite{BodlaenderDDFLP13}] \label{th:5-approx}
Let $G$ be a graph and $k$ be an integer. In time $2^{O(k)} \cdot n$, we can either decide that $\tw(G) > k$ or construct a tree-decomposition of $G$ of width at most $5k+4$.
\end{theorem}

A tree-decomposition $({\sf T}, \mathcal{B})$ rooted at a distinguished node  $t_r$ is \emph{nice} if the following conditions are fulfilled:
\begin{itemize}

\item[$\bullet$] $B_{t_r} = \es$ and this is the only empty bag,
\item[$\bullet$] each node has at most two children,
\item[$\bullet$] for each leaf $t \in V({\sf T})$, $|B_t| = 1$,
\item[$\bullet$] if $t \in V({\sf T})$ has exactly one child $t'$, then either
\begin{itemize}
\item[$\circ$] $B_t = B_{t'}\cup \{v\}$ for some $v \not \in B_{t'}$ and $t$ is called an \emph{introduce-vertex} node, or
\item[$\circ$] $B_t = B_{t'} \sm \{v\}$ for some $v \in B_{t'}$ and $t$ is called a \emph{forget-vertex} node, or
    \item[$\circ$] $B_t = B_{t'}$, $t$ is associated with an edge $\{x,y\} \in E(G)$ with $x,y \in B_t$, and $t$ is called an \emph{introduce-edge} node. We add the constraint that each edge of $G$ labels exactly one node of $T$.
\end{itemize}
\item[$\bullet$] and if $t \in V({\sf T})$ has exactly two children $t'$ and $t''$, then $B_{t} = B_{t'} = B_{t''}$. Then $t$ is called a \emph{join} node.
\end{itemize}

Note that we follow closely the definition of nice tree-decomposition given in~\cite{CyganNPPRW11}, which slightly differs from the usual one~\cite{Kloks94}. Given a tree-decomposition, then we can build a nice tree-decomposition of $G$ with the same width in polynomial time~\cite{CyganNPPRW11,Kloks94}.

Let $({\sf T}, \mathcal{B})$ be a nice tree-decomposition of a graph $G$. For each node $t \in V({\sf T}) $, we define the graph $G_t=(V_t, E_t)$ where
$V_t$ is the union of all bags corresponding to the descendant nodes of $t$, and $E_t$ is the set of all edges introduced by the descendant nodes of $t$.
Observe that the graph $G_t$ may be disconnected.


%

\paragraph{\textbf{\emph{The display graph}}.}


Let $\mathcal{T}=\{(T_1, \phi_1), (T_2, \phi_2), \ldots, (T_k, \phi_k)\}$ be a collection of unrooted phylogenetic trees. The \emph{display graph} $D_{\mathcal{T}}=(V_D, E_D)$ of $\mathcal{T}$ is the graph obtained from the disjoint union of the trees in $\mathcal{T}$ by iteratively identifying every pair of labeled vertices with the same label. We denote by $L_D$ the set of vertices of $D_{\mathcal{T}}$ resulting from these identifications.
{The elements of $L_D$ are called the \emph{labeled vertices}.}
Observe that every vertex of $V_D\setminus L_D$ (resp. every edge of $E_D$) is also a vertex (resp. an edge) of some tree $T_i\in\mathcal{T}$. If $v$ is a vertex of $L_D$, then we will say, with a slight abuse of notation, that $v$ is a vertex of $T_i$ if it results from the identification of some leaf of $T_i$. Finally, the display graph $D_{\mathcal{T}}$ is equipped with a coloring function $c: V_D \cup E_D \to \{0,\ldots,k\}$ defined as follows. If $v \in L_D$, then we set $c(v) = 0$; if $v\in (V_D\setminus L_D)\cup E_D$ belongs to the tree $T_i$, we set $c(v) = i$. Observe that if a vertex $v\in L_D$ is incident to an edge $e$ such that $c(e)=i$, then $v$ belongs to $T_i$. Suppose that $\widehat{T}$ is a (strictly) compatible supertree of $\mathcal{T}$. Then we extend the definition of vertex-model and edge-model for the vertices and edges of the $T_i$'s to the vertices and edges of the display graph $D_{\mathcal{T}}$.


The following theorem provides a bound on the treewith of the display graph of a (strictly) compatible family of unrooted phylogenetic trees:

\begin{theorem}[Bryant and Lagergren~\cite{compatibility2006}]
\label{Theo1ext}
Let $\mathcal{T}=\{(T_1, \phi_1), (T_2, \phi_2), \ldots, (T_k, \phi_k)\}$ be a collection of (strictly) compatible unrooted phylogenetic trees, not necessarily on the same label set. The display graph of $\mathcal{T}$ has treewidth at most $k$.
\end{theorem}

%

\section{Compatibility version}
\label{sec:minor}

This section provides a proof of Theorem~\ref{thm:minor}. We describe the algorithm in Subsection~\ref{sec:description}, we prove its correctness in Subsection~\ref{sec:correctness}, and we analyze its running time in Subsection~\ref{sec:runtime}.

\subsection{Description of the algorithm} 
\label{sec:description}

Let   $D=(V_D,E_D)$ be the display graph of a collection $\mathcal{T}=\{(T_1, \phi_1), (T_2, \phi_2), \ldots, (T_k, \phi_k)\}$ of unrooted phylogenetic trees, and let $n = |V(D)|$. By Theorem~\ref{th:5-approx} and Theorem~\ref{Theo1ext}, we may assume that we are given a
nice tree-decomposition $(\mathsf{T},\mathcal{B})$ of $D$ of width at most $5k+4$, as otherwise we can safely conclude that $\mathcal{T}$ is not compatible. Let  $t_r$ be the root of ${\sf T}$, and recall that $B_{t_r} = \es$.

Our objective is to build a compatible supertree $\widehat{T}$ of $\mathcal{T}$, if such exists. (We would like to note that there could exist an exponential number of compatible supertrees; we are just interested in constructing {\sl one} of them.) As it is usually the case of dynamic programming algorithms on tree-decompositions, for building $\widehat{T}$ we process $({\sf T},\mathcal{B})$ in a bottom-up way from the leaves to the root, where we will eventually decide whether a solution exists or not. We first describe the data structure used by the algorithm along with a succinct intuition behind the defined objects, and then we proceed to the description of the dynamic programming algorithm itself.



\paragraph{\textbf{\emph{Description of the data structure}}.} Before defining the dynamic-programming table associated with every node $t$ of $(\mathsf{T},\mathcal{B})$, we need a few more definitions.

\begin{definition}\label{def:supertree}
Given a node $t$ of $(\mathsf{T},\mathcal{B})$, its graph $G_t = (V_t,E_t)$, and  a subset $Z\subseteq V_t$, a  \emph{$(Z,t)$-supertree} is a tuple $\mathfrak{T}=(T,\varphi,\psi,\rho)$ such that
\begin{itemize}
\item[$\bullet$] $T$ is a tree containing at most $|B_t|+|Z|$ vertices,
\item[$\bullet$] $\varphi : Z \rightarrow 2^{V(T)}$, called the \emph{vertex-model function}, associates every $v\in Z$ with a subset $\varphi(v)$ such that
\begin{itemize}
\item[$\circ$] $T[\varphi(v)]$ is connected and if $v$ is a labeled vertex, then $|\varphi(v)|=1$, and
\item[$\circ$] if $u$ and $v$ are two vertices of $Z$ such that $c(u)=c(v)$, then $\varphi(u)\cap\varphi(v)=\emptyset$,
\end{itemize}

\item[$\bullet$] $\psi : E(T) \rightarrow 2^{[k]}$, called \emph{the edge-model function}, associates a subset of colors with every edge of $T$, and

\item[$\bullet$] $\rho: Z \to V(T)$, called the \emph{vertex-representative function}, selects, for each vertex $v \in Z$, a \emph{representative} $\rho(v)$ in the vertex-model $\varphi(v) \subseteq V(T)$.
\end{itemize}

\noindent Moreover, we say that a $(Z,t)$-supertree $(T,\varphi,\psi,\rho)$ is \emph{valid} if
\begin{itemize}
\item[$\bullet$] for every $\{u,v\} \in E_t$ such that $u, v \in Z$, then the unique edge $e$ between $\varphi(u)$ and $\varphi(v)$ exists in $T$ and satisfies $c(\{u,v\}) \in \psi(e)$.
\end{itemize}

\noindent For a node $t$ of $(\mathsf{T},\mathcal{B})$, we define a \emph{$B_t$-supertree} as a $(B_t,t)$-supertree and a \emph{$V_t$-supertree} as a $(V_t,t)$-supertree.
\end{definition}

To give some intuition on why $(Z,t)$-supertrees capture partial solutions of our problem, let us assume that $\widehat{T}$ is a compatible supertree of $\mathcal{T}$ and consider a node $t$ of $(\mathsf{T},\mathcal{B})$. Then
we can define a $B_t$-supertree $\mathfrak{T}=(T,\varphi,\psi,\rho)$ as follows:


\begin{tight_enumerate}
\item [$\bullet$] For every vertex $v \in B_t$, $\rho(v)$ can be chosen as any element in the set $\widehat{\varphi}(v)$,
\item [$\bullet$] $T=\widehat{T}|_Y$, where $Y = \underset{v \in B_t}{\bigcup} \rho(v)$,
\item [$\bullet$] for every vertex $v \in B_t$, $\varphi(v) = V(T) \cap \widehat{\varphi}(v)$,  where $\widehat{\varphi}(v)$ is the vertex-model of $v$ in $\widehat{T}$, and
\item[$\bullet$]  for every edge $e\in E(T)$,
$i\in \psi(e)$ if
there exist an edge $\{u,v\}\in E_t$, with  $c(\{u,v\})=i$, and
an edge $f\in E(\widehat{T})$
such that $f$ is incident to a vertex of $\widehat{\varphi}(u)$ and to a vertex of $\widehat{\varphi}(v)$, and
$f$ is on the unique path  in $\widehat{T}$ between the vertices incident to $e$.
\end{tight_enumerate}

\vspace{-.15cm}
The edge-model function $\psi$ introduced in Definition~\ref{def:supertree} allows to keep track, for every edge $e\in E(T)$, of the set of trees in $\mathcal{T}$ containing an edge having $e$ as an edge-model.
Observe that the size of a vertex-model $\widehat{\varphi}(v)$ in $\widehat{T}$ of some vertex $v\in V_D$ may depend on $n$ (so, a priori,  we may need to consider a number of vertex-models of size exponential in $n$). We overcome this problem via the vertex-representative function $\rho$, which allows us to store a tree $T$ of size at most $2k$. This tree $T$ captures how the vertex-models in $\widehat{T}$ ``project'' to the current bag, namely $B_t$, of the tree-decomposition of the display graph.


Before we describe the information stored at each node of the tree-decomposition, we need three more definitions.

\begin{definition}\label{def:shadow}
A tuple $\mathfrak{T}_s=(T_s,\varphi_s,\psi_s,\rho_s)$ is called a \emph{shadow} $B_t$-supertree
if there exists a $B_t$-supertree $\mathfrak{T}=(T,\varphi,\psi,\rho)$ such that
\begin{itemize}
\item[$\bullet$] $T_s$ is a tree obtained from $T$ by subdividing every edge once, called \emph{shadow tree}. The new vertices are called \emph{shadow vertices} and denoted by $S(T_s)$, while the original ones, that is, $V(T_s) \setminus S(T_s)$, are denoted by $O(T_s)$,
\item[$\bullet$] for every $v \in B_t$, $\varphi_s(v)$ is a subset of $V(T_s)$ such that $T_s[\varphi_s(v)]$ is connected and such that $\varphi(v) = \varphi_s(v) \cap O(T_s)$, where we licitly consider the vertices in $\varphi(v)$  as a subset of $O(T_s)$. Furthermore, if $u,v \in B_t$ with $c(u) = c(v)$, then $\varphi_s(u) \cap \varphi_s(v) = \emptyset$,
\item[$\bullet$] $\psi_s: E(T_s) \to 2^{\dotsss{k}}$ such that for every $s \in S(T_s)$, if $x$ and $y$ are the neighbors of $s$ in $T_s$, then $\psi_s(\{x,s\}) = \psi_s(\{s,y\}) = \psi(\{x,y\})$, and
\item[$\bullet$] $\rho_s:B_t \to V(T_s)$ such that for every $v \in B_t$, $\rho_s(v) = \rho(v)$.
\end{itemize}
\end{definition}

We say that $\TT_s$ is a \emph{shadow} of $\TT$. Note that $\TT$ may have more than one shadow satisfying Definition~\ref{def:shadow}.

\begin{definition}\label{def:restriction}
Let $\mathfrak{T}=(T,\varphi,\psi,\rho)$ be a $(Z,t)$-supertree. The \emph{restriction} of $\mathfrak{T}$ to a subset of vertices $Y\subseteq V_t$  is defined as the $(Y,t)$-supertree $\mathfrak{T}|_Y=(\tilde{T},\tilde{\varphi},\tilde{\psi},\tilde{\rho})$, where
\begin{itemize}
\item[$\bullet$]  $\tilde{T}=T|_Z$, where $Z=\{\rho(v)\mid v\in Y\}$,
\item[$\bullet$] for every $v \in Y$, $\tilde{\varphi}(v) = \varphi(v) \cap V(T|_{Y})$,
\item[$\bullet$]  for every $ e\in E(\tilde{T})$, $\tilde{\psi}(e) = \bigcup_{f \in E(P_e)} \psi(f)$, where $P_e$ is the unique path in $T$ between the vertices incident to $e$, and
\item[$\bullet$]  for every $ v \in Y$, $\tilde{\rho}(v) = \rho(v)$.
\end{itemize}
If $\TT$ is a $(Z,t)$-supertree and $B_t \subseteq Z$, we define a \emph{shadow restriction of $\TT $ to $B_t$} as a shadow of $\TT|_{B_t}$, and we denote it by
 $\TT|_{B_t}^s$.
 \end{definition}

\begin{definition}\label{def:equivalent} Two $(Z,t)$-supertrees $\mathfrak{T}=(T,\varphi, \psi, \rho)$ and $\mathfrak{T}'=(T',\varphi', \psi', \rho')$ are \emph{equivalent}, and we denote it by  $\mathfrak{T}\simeq \mathfrak{T}'$, if  there exists an isomorphism $\alpha$ from $T$ to $T'$ such that
\begin{itemize}
\item[$\bullet$]  $\forall v \in Z$, $\forall a\in \varphi(v)$, $\alpha(a)\in\varphi'(v)$,
\item[$\bullet$] $\forall e \in E(T)$, $\psi(e) = \psi'(\alpha(e))$, and 
\item[$\bullet$] $\forall v \in Z$, $\alpha(\rho(v)) = \rho'(v)$.
\end{itemize}
\end{definition}

Every  node $t$ of $(\mathsf{T},\mathcal{B})$ is associated with a set $\Rcal_t$ of pairs  $(\mathfrak{T},\gamma)$, called \emph{colored shadow $B_t$-supertrees}, where $\mathfrak{T}=(T,\varphi,\psi,\rho)$ is a shadow $B_t$-supertree and $\gamma:V(T) \rightarrow 2^{\dotsss{k}}$ is the so-called \emph{coloring function}.
The dynamic programming algorithm will maintain the following invariant:

\begin{invariant} \label{inv:dp-minor}
{A colored shadow $B_t$-supertree $(\mathfrak{T}=(T,\varphi, \psi, \rho),\gamma)$ belongs to $\Rcal_t$ if and only if there exists a valid $V_t$-supertree $\mathfrak{T}
_\ps=(T_\ps,\varphi_\ps, \psi_\ps, \rho_\ps)$ such that}

\begin{itemize}
\item[\emph{\textbf{(1)}}] $\mathfrak{T}\simeq \mathfrak{T}_\ps|_{B_t}^s$,
\item[\emph{\textbf{(2)}}] for every $a \in V(T)$, a color $i \in \gamma(a)$ if and only if there exists $ u \in V_t$ with $c(u)=i$  such that $a \in \varphi_{\ps}(u)$, and
\item[\emph{\textbf{(3)}}] for every $z \in S(T)$ with neighbors $x$ and $y$ in $V(T)$, a color $i \in \gamma(z)$ if there exists $u \in V_t$ with $c(u)=i$ and $x,y \not \in \varphi_\ps(u)$ such that
  the unique path between $x$ and $y$ in $T_\ps$ uses at least one vertex of $\varphi_\ps(u)$.
\end{itemize}
\end{invariant}

Intuitively, condition \textbf{(2)} of Invariant~\ref{inv:dp-minor} guarantees that for every vertex $v\in V(T)$, we can recover the set of trees for which $v$ has already appeared in a vertex-model of a vertex of $V_t\setminus B_t$.  On the other hand, condition \textbf{(3)} of Invariant~\ref{inv:dp-minor} is useful for the following reason. When a vertex is forgotten in the tree-decomposition, we need to keep track of its ``trace'', in the sense that the colors given to the corresponding shadow vertex guarantee that the algorithm will  construct vertex-models appropriately.
If $\gamma$ is a coloring function satisfying conditions \textbf{(2)} and \textbf{(3)}, we say that $\gamma$ is \emph{consistent} with $\mathfrak{T}_\ps$.

\smallskip

 For $Z=\emptyset$, we denote by $\oslash$ the unique colored shadow $(Z,t)$-supertree. From the above description, it follows that the collection $\mathcal{T}$ is compatible  if and only if $\oslash \in \Rcal_{t_r}$. Indeed, for $t = t_r$ we have that $B_{t_r} = \emptyset$ and $V_{t_r} = V_D$. In that case, the only condition imposed by Invariant~\ref{inv:dp-minor} is the existence of a valid $V_D$-supertree. Then, by Definition~\ref{def:supertree}, the existence of such a supertree is equivalent to the existence of a compatible supertree $\widehat{T}$ of $\mathcal{T}$ in which the vertex-models and edge-models are given by the functions $\varphi$ and $\psi$, respectively. Finally, note that the first condition of Definition~\ref{def:supertree}, namely that $|\widehat{T}| \leq |B_{t_r}| + |V_{t_r}|  = |V_D|$, is not a restriction on the set of solutions, as we may clearly assume that the size of a compatible supertree is always at most the size of the display graph.



\paragraph{\textbf{\emph{Description of the dynamic programming algorithm}}.} Let $(\mathsf{T},\mathcal{B})$ be a nice tree-decomposition of the display graph $D$ of $\mathcal{T}$. We proceed to describe how to compute the set $\Rcal_t$  for every node $t\in\mathsf{T}$. For that, we will assume inductively that, for every descendant $t'$ of $t$, we have at hand the set $\Rcal_{t'}$ that has been correctly built.  We distinguish several cases depending on the type of node $t$:

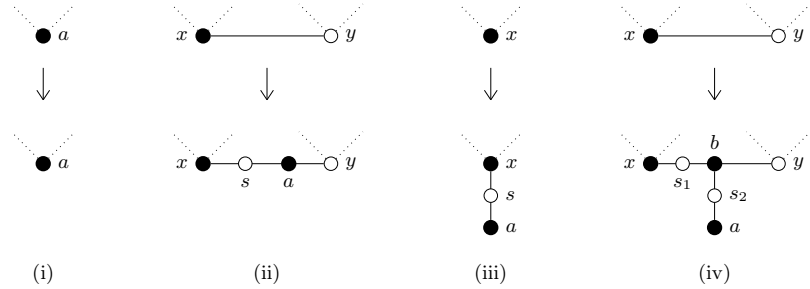
\begin{figure}[t]
\vspace{-.1cm}
\begin{center}
\vspace{.25cm}
 \scalebox{.85}{\begin{tikzpicture}

\def\x{7}
\def\y{0}

\capt{-1+\x,-4+\y}{(iii)};
\node[vertex,fill=black,label={0:$x$}, inner sep=0pt,minimum size=6pt] (x) at (\x,\y) {};
\outb{x}

\draw (0+\x,-0.5+\y) -- +(0,-0.5);
\draw (0+\x,-1+\y) -- +(0.1,0.15);
\draw (0+\x,-1+\y) -- +(-0.1,0.15);

\def\x{7}
\def\y{-2}
\node[vertex,fill=black,label={0:$x$}, inner sep=0pt,minimum size=6pt] (x) at (\x,\y) {};

\node[vertex,fill=black,label={0:$a$}, inner sep=0pt,minimum size=6pt] (a) at (\x,\y-1) {};
\draw (x) --(a);
\node[circle,draw=black,fill=white, inner sep=0pt,minimum size=6pt,label={0:$s$}] at (\x,\y-1/2) {};
\outb{x}

\def\x{9.5}
\def\y{0}
\capt{0+\x,-4+\y}{(iv)};
\node[vertex,fill=black,label={180:$x$}, inner sep=0pt,minimum size=6pt] (x) at (\x,\y) {};
\node[circle,draw=black,fill=white, inner sep=0pt,minimum size=6pt, label={0:$y$}] (y) at (\x+2,\y) {};
\draw (x) -- (y);

\outb{x}
\outb{y}

\draw (1+\x,-0.5+\y) -- +(0,-0.5);
\draw (1+\x,-1+\y) -- +(0.1,0.15);
\draw (1+\x,-1+\y) -- +(-0.1,0.15);

\def\x{9.5}
\def\y{-2}
\node[vertex,fill=black,label={180:$x$}, inner sep=0pt,minimum size=6pt] (x) at (\x,\y) {};
\node[circle,draw=black,fill=white, inner sep=0pt,minimum size=6pt, label={0:$y$}] (y) at (\x+2,\y) {};
\node[vertex,fill=black,label={90:$b$}, inner sep=0pt,minimum size=6pt] (b) at (\x+1,\y) {};
\node[vertex,fill=black,label={0:$a$}, inner sep=0pt,minimum size=6pt] (a) at (\x+1,\y-1) {};
\draw (x) -- (y);
\draw (a) -- (b);
\outb{x}
\outb{y}

\node[circle,draw=black,fill=white, inner sep=0pt,minimum size=6pt,label={-90:$s_1$}] at (\x+0.5,\y) {};
\node[circle,draw=black,fill=white, inner sep=0pt,minimum size=6pt,label={0:$s_2$}] at (\x+1,\y-0.5) {};

\def\x{2.5}
\def\y{0}
\capt{0+\x,-4+\y}{(ii)};
\node[vertex,fill=black,label={180:$x$}, inner sep=0pt,minimum size=6pt] (x) at (\x,\y) {};
\node[circle,draw=black,fill=white, inner sep=0pt,minimum size=6pt,label={0:$y$}] (y) at (\x+2,\y) {};
\draw (x) -- (y);
\outb{x}
\outb{y}


\draw (1+\x,-0.5+\y) -- +(0,-0.5);
\draw (1+\x,-1+\y) -- +(0.1,0.15);
\draw (1+\x,-1+\y) -- +(-0.1,0.15);

\def\x{2.5}
\def\y{-2}
\node[vertex,fill=black,label={180:$x$}, inner sep=0pt,minimum size=6pt] (x) at (\x,\y) {};
\node[circle,draw=black,fill=white, inner sep=0pt,minimum size=6pt,label={0:$y$}] (y) at (\x+2,\y) {};
\node[vertex,fill=black,label={-90:$a$}, inner sep=0pt,minimum size=6pt] (a) at (\x+1.34,\y) {};
\draw (x) -- (y);

\outb{x}
\outb{y}

\node[circle,draw=black,fill=white, inner sep=0pt,minimum size=6pt,label={-90:$s$}] at (\x+0.66,\y) {};

\def\x{0}
\def\y{0}
\capt{-1+\x,-4+\y}{(i)};
\node[vertex,fill=black,label={0:$a$}, inner sep=0pt,minimum size=6pt] (x) at (\x,\y) {};
\outb{x}

\draw (0+\x,-0.5+\y) -- +(0,-0.5);
\draw (0+\x,-1+\y) -- +(0.1,0.15);
\draw (0+\x,-1+\y) -- +(-0.1,0.15);

\def\x{0}

\def\y{-2}
\node[vertex,fill=black,label={0:$a$}, inner sep=0pt,minimum size=6pt] (x) at (\x,\y) {};
\outb{x}

\end{tikzpicture}}
\end{center}
\vspace{-.3cm}
\caption{The four possible cases (i-iv) in the dynamic programming algorithm. The configurations above correspond to $T'$, while the ones below correspond to $T$. Full dots correspond to vertices in $O(T)$, the other ones being in $S(T)$.\label{fig:minor}}
\end{figure}

\begin{enumerate}
\item {\bf $t$ is a leaf with $B_t=\{v\}$:} $\Rcal_{t}= \{ ((T,\varphi, \psi, \rho),\gamma) \}$, where $T$ is a tree with only one vertex $a$, $\rho(v) = a$, $\varphi(v) = \{a\}$, $\psi : \es \rightarrow 2^{\dotsss{k}}$, and $\gamma(a) = \{c(v)\}$.

\medskip
\item {\bf $t$ is an introduce-vertex node such that the introduced vertex $v$  is unlabeled:}
For every element $(\TT' = (T',\varphi',\psi',\rho'),\gamma')$ of $\Rcal_{t'}$,
we add  to $\Rcal_t$ the elements  of the form $(\TT = (T,\varphi,\psi,\rho),\gamma)$ that can be built according to one of the following four cases. For all of them, we define the vertex-representative function such that  $\rho(v)= a$ for some vertex $a \in V(T)$, and  for every $u \in B_{t'}$, $\rho(u) = \rho'(u)$. The different cases depend on this vertex $a$.

\medskip
\begin{itemize}
\item[(i)] {\bf $\rho(v) = a$ such that  $a \in V(T')$ and $c(v) \not \in \gamma'(a)$.} See Figure~\ref{fig:minor}(i) for an example.  We define $T = T'$.  Let us define  $\varphi$, $\psi$, and $\gamma$.
\begin{itemize}
\item \emph{Definition of the vertex-model function}: $T[\varphi(v)]$ is connected, contains $a$,
  and for every $z \in \varphi(v)$, $c(v) \not \in \gamma'(z)$. For every $u \in B_{t'}$, $\varphi(u) = \varphi'(u)$.
\item \emph{Definition of the edge-model function}: For every $e \in E(T)$, $\psi(e) = \psi'(e)$.
\item\emph{ Definition of the coloring function}: For every $z \in V(T)$, $\gamma(z) = \gamma'(z) \cup \{c(v) \mid  z \in \varphi(v)\}$.
\end{itemize}

\medskip
\item[(ii)] \textbf{$\rho(v)=a$  and $a$ subdivides an edge $\{x,y\}$ of $T'$ with $c(v) \not \in \psi'(\{x,y\})$}. See Figure~\ref{fig:minor}(ii) for an example.
Since $T'$ is a shadow tree, assume w.l.o.g. that $x \in O(T')$ and $y \in S(T')$.
Then $T$ is obtained from $T'$ by removing the edge $\{x,y\}$, adding two vertices $a\in O(T)$ and $s \in S(T)$ and three edges $\{x,s\}$, $\{s,a\}$, and $\{a,y\}$. Let us define  $\varphi$, $\psi$, and $\gamma$.
\begin{itemize}
\item \emph{Definition of the vertex-model function}: $T[\varphi(v)]$ is connected, contains $a$,
  and for each $z \in \varphi(v)$, $c(v) \not \in \gamma'(z)$.
  For each $u \in B_{t'}$, 
  $T[\varphi(u)]$ is connected,  $\varphi'(u) \subseteq \varphi(u) \subseteq \varphi'(u) \cup \{a\} \cup S(T)$, and
 if $u$ is unlabeled, then $\varphi(u) = \varphi'(u)$.
  For each $u,u' \in B_{t}$ with $c(u) = c(u')$, $\varphi(u) \cap \varphi(u') = \es$.

\item \emph{Definition of the edge-model function}: For each $e \in E(T) \sm \{\{x,s\}, \{s,a\},\{a,y\}\}$, $\psi(e) = \psi'(e)$. Also,
  $\psi(\{x,s\}) = \psi(\{s,a\}) = \psi(\{a,y\}) = \psi'(\{x,y\})$.

\item \emph{Definition of the coloring function}: For each $z \in O(T) \sm \{a\}$, $\gamma(z) = \gamma'(z) \cup \{c(v) \mid z \in \varphi(v)\}$.  $\gamma(a) = \{i \mid \exists u \in B_t : c(u) = i$ and $ a \in \varphi(u) \} \cup \psi'(\{x,y\})$.
For each $z \in S(T')$,   $\gamma(z) = \gamma'(z) \cup \{i \mid \exists u \in B_t : c(u) = i$ and $ z \in \varphi(u) \}$. Finally, $\gamma(s) = \{i \mid \exists u \in B_t : c(u) = i$ and $s \in \varphi(u) \} \cup \psi'(\{x,y\})$.

\end{itemize}

\medskip
\item[(iii)] \textbf{$\rho(v)=a$  with $a \notin V(T')$ and $a$ is connected to a vertex $x \in V(T')$}. See Figure~\ref{fig:minor}(iii) for an example.
 $T$ is obtained from $T'$ by adding two vertices $a\in O(T)$ and $s \in S(T)$ and two edges $\{a,s\}$ and $\{s,x\}$. Let us define $\varphi$, $\psi$, and $\gamma$.
\begin{itemize}
\item \emph{Definition of the vertex-model function}:
$T[\varphi(v)]$ is connected, contains $a$, and for each $z \in \varphi(v)$, $c(v) \not \in \gamma'(z)$.
For each $u \in B_{t'}$, 
  $T[\varphi(u)]$ is connected, $\varphi'(u) \subseteq \varphi(u) \subseteq \varphi'(u) \cup \{a\} \cup S(T)$, and
 if $u$ is unlabeled, then $\varphi(u) = \varphi'(u)$.
  For each $u,u' \in B_{t}$ with $c(u) = c(u')$, $\varphi(u) \cap \varphi(u') = \es$.

\item \emph{Definition of the edge-model function}:
For each $e \in E(T) \sm \{\{a,s\}, \{s,x\}\}$, $\psi(e) = \psi'(e)$, and
  $\psi(\{a,s\}) = \psi(\{s,x\}) = \es$.

\item \emph{Definition of the coloring function}:
  For each $z \in V(T) \sm \{a,s\}$, $\gamma(z) = \gamma'(z) \cup \{c(v) \mid z \in \varphi(v)\}$.
  For each $z \in \{a,s\}$, $\gamma(z) = \{i \mid \exists u \in B_t : c(u) = i$ and $ z \in \varphi(u) \}$.
\end{itemize}

\medskip
\item[(iv)] \textbf{$\rho(v)=a$ with $a \notin V(T')$ and $a$ subdivides an edge $\{x,y\}$ of $T'$}. See Figure~\ref{fig:minor}(iv) for an example. Again, we may assume that $x \in O(T')$ and $y \in S(T')$.
Then $T$ is obtained from $T'$ by removing the edge $\{x,y\}$,
adding four vertices $a,b \in O(T)$ and $s_1,s_2 \in S(T)$, and five edges $\{x,s_1\}$, $\{s_1,b\}$, $\{b,y\}$, $\{a,s_2\}$, and $\{s_2,b\}$. Let us define $\varphi$, $\psi$, and $\gamma$.
\begin{itemize}
\item \emph{Definition of the vertex-model function}:
$T[\varphi(v)]$ is connected, contains $a$ and, for every $z \in \varphi(v)$, $c(v) \not \in \gamma'(z)$.
For each $u \in B_{t'}$, 
  $T[\varphi(u)]$ is connected,  $\varphi'(u) \subseteq \varphi(u) \subseteq \varphi'(u) \cup \{a,b\} \cup S(T)$, and
 if $u$ is unlabeled, then $\varphi(u) = \varphi'(u)$.
For each $u,u' \in B_{t}$ with $c(u) = c(u')$, $\varphi(u) \cap \varphi(u') = \es$.

\item \emph{Definition of the edge-model function}:
For each edge $e \in E(T) \sm \{\{a,s_2\}, \{s_2,b\}, \{x,s_1\},\{s_1,b\}, \{b,y\}\}$, $\psi(e) = \psi'(e)$.
$\psi(\{x,s_1\}) = \psi(\{s_1,b\}) = \psi(\{b,y\}) = \psi'(\{x,y\})$, and
$\psi(\{a,s_2\}) = \psi(\{s_2,b\}) = \es$.

\item \emph{Definition of the coloring function}:
For every $z \in O(T) \sm \{a,b\}$, $\gamma(z) = \gamma'(z) \cup \{c(v) \mid z \in \varphi(v)\}$.
For every $z \in \{a,s_2\}$, $\gamma(z) = \{i  \mid \exists u \in B_t : c(u) = i$ and $ z \in \varphi(u) \}$.
For every $z  \in \{b,s_1\}$, $\gamma(z) = \{i \mid \exists u \in B_t : c(u) = i$ and $ z \in \varphi(u) \} \cup \psi'(\{x,y\})$.
For every $z \in S(T')$,   $\gamma(z) = \gamma'(z) \cup \{i \mid \exists u \in B_t : c(u) = i$ and $ z \in \varphi(u) \}$.
\end{itemize}
\end{itemize}

\medskip
\item {\bf $t$ is an introduce-vertex node such that the introduced vertex $v$  is labeled:} This case is very similar to Case~2 but, as vertex $v$ is a leaf, only Case~2(iii) and Case~2(iv) can be applied. In both cases, we further impose that  $\varphi(v) = \{a\}$ and $\gamma(v)=\{i\in [k]\mid v\in L(T_i), T_i\in\mathcal{T}\}$.

\medskip
\item {\bf $t$ in an introduce-edge node for an edge $\{v,w\}$ with $c(\{v,w\}) = i$:}
Let $(\TT' = (T',\varphi',\psi',\rho'),\gamma')$ be an element of $\Rcal_{t'}$ such that there exist $a \in \varphi'(v)$ and $b \in \varphi'(w)$ such that $\{a,b\} \in E(T)$ and $ i \not \in \psi'(\{a,b\})$.
We construct $(\TT = (T,\varphi,\psi,\rho),\gamma)$ as an element of $\Rcal_t$ as follows:
 $T = T'$.
For every $v \in B_t$, $\varphi(v) = \varphi'(v)$.
For every $e \in E(T) \sm \{\{a,b\}\}$, $\psi(e) = \psi'(e)$.
$\psi(\{a,b\}) = \psi'(\{a,b\}) \cup \{i\}$.
For every $v \in V(T)$, $\gamma(v) = \gamma'(v)$.

\medskip
\item
{\bf $t$ is a forget-vertex node for a vertex $v$:} Let $(\TT' = (T',\varphi',\psi',\rho'),\gamma')$ be an element of $\Rcal_{t'}$.
We construct $(\TT = (T,\varphi,\psi,\rho),\gamma)$ as an element of $\Rcal_t$ as follows: $\TT = \TT'|^z_{B_{t'}}$.
For every $a \in O(T)$, $\gamma(a) = \gamma'(a)$.
For every $z \in S(T)$, if $x$ and $y$ are the neighbors of $z$ in $T$, then $\gamma(z) = \{i \mid \exists a \in V(T')$ on the path between $x$ and $y$ in $T' : (i \in \gamma'(a))$ and $(\forall u \in B_{t}: a \not \in \varphi'(u))\}$.

\medskip
\item
{\bf $t$ is a join node:}
Let $(\TT' = (T,\varphi,\psi',\rho),\gamma')$ be an element of $\Rcal_{t'}$ and
let $(\TT'' = (T,\varphi,\psi'',\rho),\gamma'')$ be an element of $\Rcal_{t''}$ such that
for every $z \in V(T)$, $\gamma'(z) \cap \gamma''(z) = \es$ and
for every $e \in E(T)$, $\psi'(z) \cap \psi''(z) = \es$. We construct $(\TT = (T,\varphi,\psi,\rho),\gamma)$ as an element of $\Rcal_t$ as follows:
For every $e \in E(T)$, $\psi(e) = \psi'(e) \cup \psi''(e)$, and
 for every $z \in V(T)$, $\gamma(z) = \gamma'(z) \cup \gamma''(z)$.

\end{enumerate}


\subsection{Correctness of the algorithm}
\label{sec:correctness}
Let $t$ be a node of $(\mathsf{T},\mathcal{B})$. Our objective is to prove that, on the one hand, the elements $(\TT,\gamma)$ generated by the algorithm indeed belong to the set $\Rcal_t$ (that is, that they satisfy Invariant~\ref{inv:dp-minor}) and, on the other hand, that all the elements of the set $\Rcal_t$ are constructed by the algorithm.  We will assume inductively that both claims are true for every descendant $t'$ of $t$. 

Our approach for proving that the generated elements belong to $\Rcal_t$ is the following. We distinguish again the cases of the algorithm. For each of them, the assumption that  $\Rcal_{t'}$ has been correctly built for every descendant $t'$ of $t$ guarantees the existence, for every element $(\TT',\gamma')$ of $\Rcal_{t'}$, of the corresponding  certificate $\TT'_\ps$ that implies by Invariant~\ref{inv:dp-minor} that $(\TT',\gamma') \in \Rcal_{t'}$. We will then use $\TT'_\ps$ to prove, for each of the elements $(\TT,\gamma)$ constructed by the algorithm, that there {\sl exists} a certificate $\TT_\ps$ implying that $(\TT,\gamma) \in \Rcal_{t}$.

We would like to stress that, in order to prove that $(\TT,\gamma) \in \Rcal_{t}$, we only need to worry about the {\sl existence} of such a certificate $\TT_\ps$, and not about how it can be {\sl constructed}. However, if we are interested in constructing a compatible supertree (and not only knowing whether it exists or not), we can easily do it as well. Indeed, starting from the leaves of the tree-decomposition, by using the operations described below we can inductively grow the certificates $\TT'_\ps$ of $\Rcal_{t'}$ to get the certificates $\TT_\ps$ of $\Rcal_{t}$, within the same running time of the algorithm.

We now proceed to distinguish the different cases of the dynamic programming algorithm presented in Subsection~\ref{sec:description}:

\begin{enumerate}
\item {\bf $t$ is a leaf with $B_t=\{v\}$:}
$T_\ps$ is a tree with only one vertex $a$, $\rho_\ps(v) = a$, $\varphi_\ps(v) = \{a\}$, and $\psi_\ps : \es \rightarrow 2^{\dotsss{k}}$.

\medskip
\item {\bf $t$ is an introduce-vertex node such that the introduced vertex $v$  is unlabeled:}
 Given an element $(\TT', \gamma')$ of $\Rcal_{t'}$ with the corresponding certificate $\TT'_\ps$, we distinguish the different cases of the algorithm that create  elements of the form $(\TT, \gamma)$, and we define for each case a certificate $\TT_\ps$ of $\TT$, which implies that $(\TT, \gamma) \in \Rcal_t$.

\medskip
\begin{itemize}
\item[(i)] {\bf $\rho(v) = a$ such that  $a \in V(T')$ and $c(v) \not \in \gamma'(a)$.}
Then $T_\ps = T'_\ps$.
Let us define $\rho_\ps$, $\varphi_\ps$, and $\psi_\ps$.
\begin{itemize}
\item \emph{Definition of the vertex-representative function}:
\begin{itemize}
\item
  $\rho_\ps(v) = \rho(v)= a$ and
\item
  for every $u \in V_{t'}$, $\rho_\ps(u) = \rho'_\ps(u)$.

\end{itemize}
\item\emph{ Definition of the vertex-model function}:
\begin{itemize}
 \item $T_\ps[\varphi_\ps(v)]$ is connected and contains $a$,
$\varphi(v) \cap O(T) = \varphi_\ps(v) \cap O(T)$,
\item
  for every $u \in B_{t'}$, $\varphi_\ps(u) = \varphi'_\ps(u)$, and
\item
  for every $u,u' \in V_{t}$ with $c(u) = c(u')$, $\varphi_\ps(u) \cap \varphi_\ps(u') = \es$.

\end{itemize}
\item \emph{Definition of the edge-model function}:
\begin{itemize}
\item  for every $e \in E(T)$, $\psi_\ps(e) = \psi'_\ps(e)$.

\end{itemize}
\end{itemize}

\medskip
\item[(ii)] \textbf{$\rho(v)=a$  and $a$ subdivides an edge $\{x,y\}$ of $T'$ with $c(v) \not \in \psi'(\{x,y\})$}.
$T_\ps$ is obtained from $T'_\ps$ by removing an edge $\{x_\ps,y_\ps\}$ on the path between $x$ and $y$, and adding a vertex $a$ and two edges $\{x_\ps,a\}$ and $\{a,y_\ps\}$.
Let us define $\rho_\ps$, $\varphi_\ps$, and  $\psi_\ps$.
\begin{itemize}
\item \emph{Definition of the vertex-representative function}:
\begin{itemize}
\item
  $\rho_\ps(v) = \rho(v) = a$ and
\item
  for every $u \in V_{t'}$, $\rho_\ps(u) = \rho'_\ps(u)$.
\end{itemize}
\item \emph{Definition of the vertex-model function}:
\begin{itemize}
\item
  $T_\ps[\varphi_\ps(v)]$ is connected and contains $a$,
\item
  for every $u \in V_{t'}$,
  $T_\ps[\varphi_\ps(u)]$ is connected, $\varphi'_\ps(u) \subseteq \varphi_\ps(u) \subseteq \varphi'_\ps(u) \cup \{a\}$, and
 if $u$ is unlabeled, then $\varphi_\ps(u) = \varphi'_\ps(u)$,
\item
  for every $u \in B_{t}$,   $\varphi(u) \cap O(T) = \varphi_\ps(u) \cap O(T)$,
\item
  for every $u,u' \in V_{t}$ with $c(u) = c(u')$, $\varphi_\ps(u) \cap \varphi_\ps(u') = \es$, and
\item
  for every $\{u,u'\} \in E_t$, there exist $w \in \varphi_\ps(u)$ and $w' \in \varphi_\ps(u')$ such that $\{w,w'\} \in E(T_\ps)$.
\end{itemize}
\item \emph{Definition of the edge-model function}:
\begin{itemize}

\item
  for every $e \in E(T) \sm \{\{x_\ps,a\},\{a,y_\ps\}\}$, $\psi_\ps(e) = \psi'_\ps(e)$ and
\item
  $\psi_\ps(\{x_\ps,a\}) = \psi_\ps(\{a,y_\ps\}) = \psi'_\ps(\{x_\ps,y_\ps\})$.
\end{itemize}
\end{itemize}

\medskip
\item[(iii)] \textbf{$\rho(v)=a$  with $a \notin V(T')$ and $a$ is connected to a vertex $x \in V(T')$}.
$T_\ps$ is obtained from $T'_\ps$ by adding a vertex $a$  and an edge $\{a,x\}$.
Let us define $\rho_\ps$, $\varphi_\ps$, and $\psi_\ps$.
\begin{itemize}
\item \emph{Definition of the vertex-representative function}:
\begin{itemize}
\item
  $\rho_\ps(v) = \rho(v) = a$ and
\item
  for every $u \in V_{t'}$, $\rho_\ps(u) = \rho'_\ps(u)$.
\end{itemize}
\item \emph{Definition of the vertex-model function}:
\begin{itemize}
\item
  $T_\ps[\varphi_\ps(v)]$ is connected and contains $a$,
\item
  for every $u \in V_{t'}$, 
  $T_\ps[\varphi_\ps(u)]$ is connected, $\varphi'_\ps(u) \subseteq \varphi_\ps(u) \subseteq \varphi'_\ps(u) \cup \{a\}$, and
 if $u$ is unlabeled, then $\varphi_\ps(u) = \varphi'_\ps(u)$,

\item
  for every $u \in B_{t}$,   $\varphi(u) \cap O(T) = \varphi_\ps(u) \cap O(T)$,
\item
  for every $u,u' \in V_{t}$ with $c(u) = c(u')$, $\varphi_\ps(u) \cap \varphi_\ps(u') = \es$, and
\item
  for every $\{u,u'\} \in E_t$, there exist $w \in \varphi_\ps(u)$ and $w' \in \varphi_\ps(u')$ such that $\{w,w'\} \in E(T_\ps)$.
\end{itemize}
\item \emph{Definition of the edge-model function}:
\begin{itemize}
\item
  for every $e \in E(T) \sm \{\{a,x\}\}$, $\psi_\ps(e) = \psi'_\ps(e)$ and
\item
  $\psi(\{a,x\}) = \es$.
\end{itemize}
\end{itemize}

\medskip
\item[(iv)] \textbf{$\rho(v)=a$ with $a \notin V(T')$ and $b$ subdivides an edge $\{x,y\}$ of $T'$}.
$T_\ps$ is obtained from $T'_\ps$ by removing an edge $\{x_\ps,y_\ps\}$ on the path between $x$ and $y$, and adding two vertices $a$ and $b$ and three edges $\{x_\ps,b\}$, $\{b,y_\ps\}$, and $\{a,b\}$.
Let us define $\rho_\ps$, $\varphi_\ps$, and  $\psi_\ps$.

\begin{itemize}
\item \emph{Definition of the vertex-representative function}:
\begin{itemize}
\item
  $\rho_\ps(v) = \rho(v) = a$ and
\item
  for  every $u \in V_{t'}$, $\rho_\ps(u) = \rho'_\ps(u)$.
\end{itemize}
\item \emph{Definition of the vertex-model function}:
\begin{itemize}
\item

  $T_\ps[\varphi_\ps(v)]$ is connected and contains $a$,
\item
  for every $u \in V_{t'}$, 
  $T_\ps[\varphi_\ps(u)]$ is connected, $\varphi'_\ps(u) \subseteq \varphi_\ps(u) \subseteq \varphi'_\ps(u) \cup \{a,b\}$, and
 if $u$ is unlabeled, then $\varphi_\ps(u) = \varphi'_\ps(u)$,

\item
  for every $u \in B_{t}$,   $\varphi(u) \cap O(T) = \varphi_\ps(u) \cap O(T)$,

\item
  for every $u,u' \in V_{t}$ with $c(u) = c(u')$, $\varphi_\ps(u) \cap \varphi_\ps(u') = \es$, and
\item
  for every $\{u,u'\} \in E_t$, there exist $w \in \varphi_\ps(u)$ and $w' \in \varphi_\ps(u')$ such that $\{w,w'\} \in E(T_\ps)$.
\end{itemize}
\item \emph{Definition of the edge-model function}:
\begin{itemize}

\item

  for every $e \in E(T) \sm \{\{a,b\}, \{x_\ps,b\}, \{b,y_\ps\}\}$, $\psi(e) = \psi'(e)$,
\item
  $\psi_\ps(\{x_\ps,b\}) = \psi_\ps(\{b,y_\ps\}) = \psi_\ps'(\{x_\ps,y_\ps\})$, and
\item
  $\psi_\ps(\{a,b\}) = \es$.
\end{itemize}
\end{itemize}
\end{itemize}

\medskip
\item {\bf $t$ is an introduce-vertex node such that the introduced vertex $v$  is labeled:}
As explained in the description of the algorithm, this case is very similar to Case~2, taking into account that only Case~2(iii) and Case~2(iv) can be applied, and by adding the following constraints:

\begin{itemize}
\item[$\bullet$] $\varphi_\ps(v) = \{a\}$ and
\item[$\bullet$] $\gamma_\ps(v)=\{i\in [k]\mid v\in L(T_i), T_i\in\mathcal{T}\}$.
\end{itemize}
\end{enumerate}

\smallskip

\noindent In the next two cases, let $(\TT', \gamma')$ be the element of $\Rcal_{t'}$ from which the algorithm has started, let $\TT'_\ps$ be a certificate of $(\TT', \gamma')$, and let $(\TT, \gamma)$ be the element created by the algorithm. In both cases, we construct  a certificate $\TT_\ps$ of $(\TT, \gamma')$ showing that $(\TT, \gamma) \in \Rcal_t$.

\smallskip
\begin{enumerate}
\item[4.] {\bf $t$ in an introduce-edge node for an edge $\{v,w\}$ with $c(\{v,w\}) = i$:} We construct $\TT_\ps = (T_\ps,\varphi_\ps,\psi_\ps,\rho_\ps)$ as follows:
\begin{itemize}
\item[$\bullet$] $T_\ps = T'_\ps$,
\item[$\bullet$] for every $v \in V_t$, $\varphi_\ps(v) = \varphi'_\ps(v)$,
\item[$\bullet$] for every $e \in E(T) \sm \{\{a,b\}\}$, $\psi_\ps(e) = \psi'_\ps(e)$, and
\item[$\bullet$] $\psi_\ps(\{a,b\}) = \psi_\ps'(\{a,b\}) \cup \{i\}$.

\end{itemize}

\medskip
\item[5.]
{\bf $t$ is a forget-vertex node for a vertex $v$:} In this case, we just define $\TT_\ps = \TT'_\ps$.


\medskip
\item[6.]
{\bf $t$ is a join node:}
Let $(\TT', \gamma')$ be the element of $\Rcal_{t'}$ and  let $(\TT'', \gamma'')$ be the element of $\Rcal_{t''}$ from which the algorithm has started, and let $\TT'_\ps= (T_\ps,\varphi_\ps,\psi'_\ps,\rho_\ps)$ and $\TT''_\ps=(T_\ps,\varphi_\ps,\psi''_\ps,\rho_\ps)$ be their certificates, respectively. We define $\TT_\ps =  (T_\ps,\varphi_\ps,\psi_\ps,\rho_\ps)$, that is, a certificate of $(\TT, \gamma)$  showing that $(\TT  , \gamma) \in \Rcal_t$, just by setting, for every $e \in E(T)$, $\psi_\ps(e) = \psi'_\ps(e) \cup \psi''_\ps(e)$. Note that $T_\ps$, $\varphi_\ps$, and $\rho_\ps$ are those given by $(\TT', \gamma')$ (or by $(\TT'', \gamma'')$).



\end{enumerate}

\bigskip

Finally, let us argue that all the elements of the set $\Rcal_t$ are indeed constructed by the algorithm. Let $(\TT,\gamma)$ be an element of $\Rcal_t$, with $\TT = (T, \varphi, \psi, \rho)$, and our objective  is to show that the algorithm indeed generates this element $(\TT,\gamma)$.
In order to do this, we need to consider each case of the algorithm separately. We will only detail the arguments for  Case 2, which is the most involved one, and the other ones follow by using a similar argumentation.

By definition of the set $\Rcal_t$, there exists a valid $V_t$-supertree $\TT_{\ps}$ such that $\TT = \TT_{\ps}|_{B_t}^s$ and such that $\gamma$ is consistent with $\TT_{\ps}$.
Let $\TT_{\ps}' = \TT_{\ps}|_{V_{t'}}$.
It can be easily checked that $\TT_{\ps}'$ is a valid $V_{t'}$-supertree.
Let $\TT' = \TT_{\ps}'|_{B_{t'}}^s$ and let $\gamma'$ be the coloring function consistent with $\TT_{\ps}'$.
Then, as Invariant~\ref{inv:dp-minor} is satisfied, $(\TT',\gamma')$ is an element of $\Rcal_{t'}$.
Note that $\TT'  = \TT|_{B_{t'}}$.
As the sets $B_{t}$ and $B_{t'}$ differ by just one vertex, the elements $\TT$ and $\TT'$ are quite close to each other. Indeed, the way they differ is mainly given  by the value of $\rho(v)$, in the sense that we consider all the possible ways  to add a vertex $\rho(v)$ to a tree $T'$.
It appears that there are four different ways to add $\rho(v)$ to $T'$.
Indeed, $\rho(v)$ can either be an already existing vertex of $T'$, or a new vertex that subdivides an edge, or a new vertex connected to an already existing vertex, or a new vertex connected to another new vertex that subdivides an edge.
Our algorithm precisely explore these four possibilities for $\rho(v)$, and then updates $T$, $\varphi$, $\psi$, and $\gamma$ in all the possible ways such that the resulting element is still in $\Rcal$.
So in particular, the algorithm necessarily created the element $(\TT,\gamma)$ of $\Rcal_t$, as we wanted to show.

\subsection{Running time analysis of the algorithm}
\label{sec:runtime}
Let us now discuss the running time of the dynamic programming algorithm described in Subsection~\ref{sec:description}. Let $w$ be the width of  $({\sf T},\mathcal{B})$, so we have that $w \leq 5k + 4$. For each $t \in V({\sf T})$, we bound the size of $\Rcal_t$ as follows. Each element in $\Rcal_t$ is of the form $(\TT =(T,\varphi, \psi, \rho), \gamma)$. Note that $T$ has at most $3 w$ nodes, and that there are at most $(3w)^{3w-2} = 2^{\Ocal(k \log k)}$ distinct trees on $3w$ vertices~\cite{Cay89}. There are at most $2^{|V(T)|\cdot|B_t|} \leq 2^{3w \cdot w}$ possible functions $\varphi$, $2^{|E(T)|\cdot k} \leq 2^{3w \cdot k}$ possible functions $\psi$, $|V(T)|^{|B_t|} \leq (3w)^{w}$ possible functions $\rho$, and $2^{|V(T)|\cdot k} \leq 2^{3w \cdot k}$ possible functions $\gamma$.
Thus, it holds that $|\Rcal_t| = 2^{\Ocal(k^2)}$ for every node $t$ of $({\sf T},\mathcal{B})$.

Concerning the complexity of computing $\Rcal_t$, we distinguish several cases. This computation is trivial in Case~1 of the algorithm, that is, when $t$ is a leaf. In Cases 2, 3, 4, and 5, the set $\Rcal_t$ can be clearly computed in time polynomial in $|\Rcal_{t'}|$, where $t'$ is the child of $t$. Finally, in Case~6, that is, when $t$ is a join node, the set $\Rcal_t$ can also be clearly computed in time polynomial in $|\Rcal_{t'}|$ and $|\Rcal_{t''}|$, where $t'$ and $t''$ are the two children of $t$. Finally, as we can assume that  $|V({\sf T})| = \Ocal(n)$~\cite{Kloks94},  the running time claimed in Theorem~\ref{thm:minor} follows.

\section{Agreement version}
\label{sec:topological}
In this section we provide a proof of Theorem~\ref{thm:topo}. Again,  by Theorem~\ref{th:5-approx} and Theorem~\ref{Theo1ext}, we may assume that we are given a nice tree-decomposition $(\mathsf{T},\mathcal{B})$ of $D$ of width at most $5k+4$.

The algorithm follows closely the one described in Subsection~\ref{sec:description} for the compatibility version, so we will just describe the changes to be done to deal with the agreement version. Intuitively, these changes appear because  now we are looking for a supertree containing each of the trees in $\mathcal{T}$ as a \emph{topological minor}, instead of a \emph{minor}, and this forces us to redefine the notions of vertex-model and edge-model functions. Namely, each vertex-model becomes a \emph{single vertex} (instead of a set of vertices), and for guaranteeing the existence of the appropriate topological minors, we have to keep track of the existence of pairwise disjoint \emph{paths} among the vertex-models of each color (instead of just edges).

We first proceed to partially redefine the data structure, and then we will focus on the changes in the dynamic programming algorithm.

\paragraph{\textbf{\emph{Changes in the data structure}}.} For a node $t$ of the tree-decomposition, our tables  $\Rcal_t$ store again elements of the form $(\mathfrak{T},\gamma)$ satisfying the same invariant as in Subsection~\ref{sec:description}, namely Invariant~\ref{inv:dp-minor}, the difference is that we update some definitions of the data structure. Namely, the vertex-model function in the definition of $(Z,t)$-supertree, cf. Definition~\ref{def:supertree}, is updated as follows:

\begin{itemize}
\item[$\bullet$] $\varphi : Z \rightarrow V(T)$ is such that if $u$ and $v$ are two vertices of $Z$ with $c(u)=c(v)$, then $\varphi(u) \neq \varphi(v)$,
\end{itemize}


\noindent We also modify slightly  the definition of ``valid supertrees'' and say that a $(Z,t)$-supertree $(T,\varphi,\psi,\rho)$ is \emph{valid} if
\begin{itemize}
\item[$\bullet$] for every $\{u,v\} \in E_t$ such that $u, v \in Z$, every edge $e$ on the path between $\varphi(u)$ and $\varphi(v)$  in $T$ satisfies $c(\{u,v\}) \in \psi(e)$ and
\item [$\bullet$] if $i \in \psi(e)$ for some $i \in \dotsss{k}$, then there exists a unique pair $\{u,v\} \in E_t$ with $u,v \in Z$ with $c(\{u,v\}) = i$ such that $e$ lies on the path between $\varphi(u)$ and $\varphi(v)$.
\end{itemize}



It is worth noting that the dynamic programming algorithm described below satisfies that, for every vertex $v \in Z$, $\varphi(v) = \rho(v)$, and therefore the vertex-representative function $\rho$ becomes superfluous. Nevertheless, in order for the notation to deviate as little as possible to that of Section~\ref{sec:minor}, we keep $\rho$ in the tuple $\mathfrak{T}$.

\paragraph{\textbf{\emph{Changes in the dynamic programming algorithm}}.} The fact that the image of the vertex-model function $\varphi$ is now a single vertex allows us to substantially simplify the algorithm.
In particular, in the subcases of the two cases where $t$ is an introduce-vertex node (namely, Cases 2 and 3),
 we do not have to worry anymore about how the image of $\varphi$ grows when introducing a new vertex, except, naturally,  for this newly introduced vertex. The latter simplification implies that we do not need to update the coloring function $\gamma$ either, except again for the newly introduced vertex. Finally, as the function $\rho$ is now redundant, we may omit it from the description of the algorithm.

More precisely,  Cases 1, 2, 3, 5, and 6 of the algorithm from Subsection~\ref{sec:description} remain unchanged, just by taking into account that $\varphi(v)$ returns just one element, namely $\varphi(v)=a$.
The changes occur in Case 4, which becomes as follows:
\begin{itemize}
\item[4.] {\bf $t$ in an introduce-edge node for an edge $\{v,w\}$ with $c(\{v,w\}) = i$:}
Let $(\TT' = (T',\varphi',\psi',\rho'),\gamma')$ be an element of $\Rcal_{t'}$
such that
for each $e \in P_{v,w}$, $i \not \in \psi'(e)$, where $P_{v,w} = \{e \in E(T) \mid e$ lies on the path between $\varphi(v)$ and $\varphi(w)\}$.
We construct $(\TT = (T,\varphi,\psi,\rho),\gamma)$ as an element of $\Rcal_t$ as follows:
\begin{itemize}
\item[$\bullet$] $T = T'$,
\item[$\bullet$] for every $v \in B_t$, $\varphi(v) = \varphi'(v)$,
\item[$\bullet$] for every $e \in E(T) \sm P_{v,w}$, $\psi(e) = \psi'(e)$,
\item[$\bullet$] for every $e \in P_{v,w}$, $\psi(e) = \psi'(e) \cup \{i\}$, and
\item[$\bullet$] for every $v \in V(T)$, $\gamma(v) = \gamma'(v)$.
\end{itemize}
\end{itemize}

\bigskip

The correctness of the algorithm can be proved analogously to the proof given in Subsection~\ref{sec:correctness}. Finally, note that the analysis of the running time carried out in Subsection~\ref{sec:runtime} also applies to this case, as the size of the objects stored in the tables is upper-bounded by the size of those used in the algorithm of Subsection~\ref{sec:description}. Furthermore, the performed operations incur the same time complexity, except for the case of an introduce-edge node, for which in the previous algorithm we looked for the existence of an appropriate {\sl edge} in $T$, whereas in the current one we look for the existence of an appropriate {\sl path} in $T$, which can be performed in time $O(|V(T)|)$. This additional running time is clearly dominated by the overall running time of the algorithm, namely $2^{O(k^2)} \cdot n$.



\section{Further research}
\label{sec:conclusions}

In this paper we give the first ``reasonable'' FPT algorithms for the {\sc Compatibility} and the {\sc Agreement} problems for  unrooted phylogenetic trees.
Even though this is, from a theoretical point of view, a big step further toward solving this problem in reasonable time, our running times are still prohibitive to be of any use in real-life phylogenomic studies,  where $k$ can go up very quickly  \cite{delsuc2005phylogenomics}. One possibility to design a practical algorithm is to devise reduction rules to keep $k$ small. Another possibility would be to design an FPT algorithm with respect to a parameter that  is smaller than  the number of gene trees in phylogenomic studies.

From a more theoretical perspective, a natural question is whether the function $2^{O(k^2)}$ in the running times of our algorithms can be improved. It would also be interesting to prove lower bounds for algorithms parameterized by treewidth to solve these problems, assuming the Exponential Time Hypothesis~\cite{LokshtanovMS11}.


{\small
\bibliographystyle{abbrv}
\bibliography{biblio}
}

\end{document}